\documentclass[11pt,aps,pra,reprint,superscriptaddress]{revtex4-2}

\usepackage{graphicx}
\usepackage{hyperref}
\usepackage{braket}
\usepackage{amsmath}
\usepackage{amssymb}
\usepackage{mathtools}
\usepackage{dsfont}
\usepackage{systeme}
\usepackage{titlesec}
\usepackage{bbm}
\usepackage{xcolor}
\DeclareMathOperator{\sech}{sech}

\begin{document}


\title{Quantum kernels with squeezed-state encoding for machine learning}


\author{Long Hin Li}
\email{lhliac@connect.hku.hk}
\affiliation{Guangdong-Hong Kong Joint Laboratory of Quantum Matter, Department of Physics, and HKU-UCAS Joint Institute for Theoretical and Computational Physics at Hong Kong, The University of Hong Kong, Pokfulam Road, Hong Kong, China}

\author{Dan-Bo Zhang}
\email{dbzhang@m.scnu.edu.cn}
\affiliation{Guangdong-Hong Kong Joint Laboratory of Quantum Matter, Frontier Research Institute for Physics,
South China Normal University, Guangzhou 510006, China}
\affiliation{Guangdong Provincial Key Laboratory of Quantum Engineering and Quantum Materials, School of Physics and Telecommunication Engineering, South China Normal University, Guangzhou 510006, China}

\author{Z. D. Wang}
\email{zwang@hku.hk}
\affiliation{Guangdong-Hong Kong Joint Laboratory of Quantum Matter, Department of Physics, and HKU-UCAS Joint Institute for Theoretical and Computational Physics at Hong Kong, The University of Hong Kong, Pokfulam Road, Hong Kong, China}

\date{\today}

\begin{abstract}
Kernel methods are powerful for machine learning, as they can represent data in feature spaces that similarities between samples may be faithfully captured. Recently, it is realized that machine learning enhanced by quantum computing is closely related to kernel methods, where the exponentially large Hilbert space turns to be a feature space more expressive than classical ones. In this paper, we generalize quantum kernel methods by encoding data into continuous-variable quantum states, which can benefit from the infinite-dimensional Hilbert space of continuous variables. Specially, we propose squeezed-state encoding, in which data is encoded as either in the amplitude or the phase. The kernels can be calculated on a quantum computer and then are combined with classical machine learning, e.g. support vector machine, for training and predicting tasks. Their comparisons with other classical kernels are also addressed. Lastly, we discuss physical implementations of squeezed-state encoding for machine learning in quantum platforms such as trapped ions.

\end{abstract}

\maketitle

\section{Introduction}
Machine learning can be enhanced by quantum computing utilizing the incredible power of quantum computers for information processing~\cite{biamonte_17,lloyd_14,rebentrost_14,dunjko_16,lloyd_16,lloyd_18,havlicek_19,schuld_19,huang2021power,liu_rigorous_2021}. The first breakthrough comes from an exponential quantum speedup for solving linear equations or matrix inversion~\cite{harrow_09,wiebe_12}, which lies at the heart of many machine learning methods. While quantum computers are good at linear algebra since they in nature perform unitary transformations that are linear in nature, it is not enough for machine learning of complicated data, where highly nonlinear transformations are often anticipated~\cite{Goodfellow-et-al-2016}. Another important ingredient for quantum-enhanced machine learning, only recognized very recently, is to introduce non-linearity when encoding classical data into quantum states~\cite{mitarai_18,havlicek_19,schuld_19,lloyd2020quantum,PerezSalinas2020datareuploading,Schuld-2021-PRA,schuld2021supervised}. Such a procedure is called quantum feature maps~\cite{havlicek_19,schuld_19}, which enjoy the exponential large Hilbert space as a feature space for expressing data.

The quantum feature maps are closely related to classical kernel methods in the same spirit as they aim to use feature spaces, such that similarities between samples can be faithfully captured~\cite{christopher_m_bishop_pattern_2006,trevor_hastie_elements_2009}. One striking difference is that feature maps on a quantum computer are explicitly constructed~\cite{schuld2021supervised}. This gives the flexibility of designing or even training powerful quantum kernels that may provide quantum advantages for machine learning. Varies encoding schemes for quantum kernels have been explored. Typically, data vectors are encoded as parameters of quantum gates, such as rotation angles of single-qubit gates. By loading the same data repeatedly into different quantum gates, the quantum data state will be a nonlinear function of the data, which consists of both different high-frequency and low-frequency components~\cite{Schuld-2021-PRA}. Those types of quantum kernels are rather different from classical ones, and while useful applications can be expected, they would have limitations(e.g., similarity between two samples may not be nicely approximated as a periodic function of data vector). On the other hand, quantum computing and quantum machine learning with continuous variables can inherit many properties of classical machine learning~\cite{lau_17,zhang_19,schuld_19,Killoran-cv-NN-2019,zhang-2020-PRL}, e.g., the widely-applied Gaussian kernel can be realized on a quantum computer by coherent-state encoding. Moreover, squeezed-state encoding has been proposed in Ref.~\cite{schuld_19}. This suggests that quantum feature maps by encoding data as continuous-variable states can be promising, by exploiting the infinite dimensionality of continuous-variable for encoding high-density information. 

In this paper, we focus on one important class of continuous-variable states, the squeezed state, as a quantum kernel for machine learning. We first introduce the squeezing vacuum state, and two different encoding schemes, the amplitude encoding and the phase encoding. Those different data encoding methods can give different kernels, which have their unique properties for modeling data. The kernel can be calculated as an inner product between two quantum states that encode the two data, which further serves as input for machine learning models, such as support vector machines.  We will then test those kernels and also compare them with some classical kernels on different datasets classification tasks. Finally, we propose a physical platform for implementation and also discuss further possibilities to expand the kernel libraries.


\section{Quantum kernels with squeezed-state encoding}
\label{sec:kernels}
Let us first introduce some notations. For a sample represented by a data vector $x$, its quantum feature map is realized by encoding the data into a quantum state $\ket{\psi_x}$. The quantum kernel of two samples $x$ and $x'$
is an inner product between their corresponding quantum states, namely $K(x,x')=\braket{\psi_x|\psi_{ x'}}$. 

A squeezed vacuum state is a minimum uncertainty state satisfying $\Delta\hat{x}\Delta\hat{p}=\frac{1}{4}$, where $\hat{x}$ and $\hat{p}$ and conjugate quadrature operators with either quadrature variance squeezed~\cite{weedbrook_12}. It can be obtained from the vacuum state $\ket{0}$,
\begin{align}
\ket{z}_s=\hat{S}(z)\ket{0}
\end{align}
with squeezing operator defined as $\hat{S}(z)=e^{\frac{1}{2}(z^*\hat{a}^2-z\hat{a}^{\dagger 2})}$. The position uncertainty is "squeezed" if the amplitude $r$ of the squeezing parameter $z=re^{i\theta}$ is positive. The momentum uncertainty will be enlarged by the same proportion to keep their product constant. The squeezed state can be expanded in terms of Fock state basis by using the property $\hat{S}(z)\hat{a}\hat{S}^\dagger (z)=\hat{a}\cosh{r}+\hat{a}^\dagger e^{i\theta} \sinh{r}$,
\begin{align}
\ket{z}_s=\frac{1}{\sqrt{\cosh{r}}} \sum_{n=0}^\infty (-1)^n \frac{\sqrt{(2n)!}}{2^n n!}e^{in\theta}\tanh^n{r} \ket{2n} \label{ss}
\end{align}

Here we introduce the notation separating the amplitude and phase of the squeezed vacuum state $\ket{z}=\ket{r,\theta}$.
The inner product between squeezed vacuum states is
\begin{align}
\braket{r,\theta|r',\theta'}_s=\sqrt{\frac{\sech{r}\sech{r'}}{1-e^{i(\theta'-\theta)}\tanh{r}\tanh{r'}}} \label{spk}
\end{align}

The inner product will serve as a measure of similarity between data, equivalent to the kernel in machine learning~\cite{trevor_hastie_elements_2009}. This provides a bridge from quantum computing to machine learning. In many instances, we use basic encoding $\ket{\psi_x}=\ket{x}$ with $\ket{x}$ a tensor product of binary qubits or amplitude encoding $\ket{\psi_x}=\sum_{n=0}^N x_n\ket{n}$ to encode data in quantum algorithms. Apart from the discrete variable qubit, encoding with coherent states are also possible~\cite{zhang-2020-PRL}. Different feature maps are able to capture different data patterns. And with the greater amount of possible feature maps, more data patterns can be approximated. For example, the amplitude encoding gives the usual dot product and basic encoding on coherent state gives Gaussian kernel.

\subsection{Phase encoding}

\begin{figure}[h!]
	\begin{center}
		\includegraphics[width=0.5\textwidth]{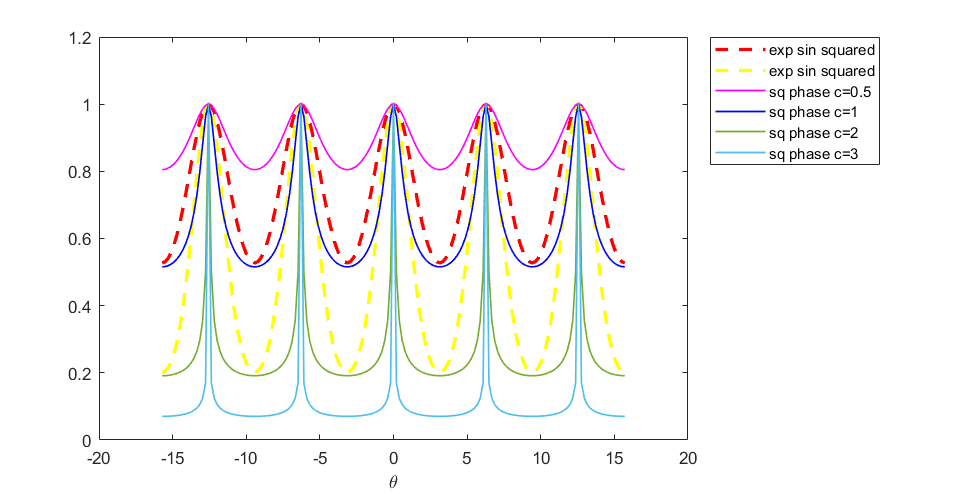}
		\caption{Comparison between exponential sine squared kernel for different $l$ from Eq. \ref{ess} (dash line) and squeezing phase kernel for different $c$ from Eq. \ref{spk} (solid line).} \label{argu}
	\end{center}
\end{figure}

Given an $N$-dimensional dataset $\{{\bf x}_m\}$, we can encode data into the phase parameters of $N$ squeezed vacuum modes with the amplitude as a hyperparameter~\cite{schuld_19}. The encoding was $\phi: x \rightarrow \ket{c,x}_s$, with $\ket{c,x}_s$ defined in Eq.~(\ref{ss}) and is referred as squeezing feature map with phase encoding. We call this kernel the {\it squeezing phase kernel}. The kernel plot for different data entries is a periodic function, as shown in Fig.~\ref{argu}. The similarity between data valued at multiples of $2\pi$ separations is high. This shares similar properties to the exponential sine squared kernel in machine learning

\begin{align}
K_{ess}(x,x') = \exp \left(-\frac{2}{l^2}\sin^2 \left( |x-x'|\pi/p \right)\right) \label{ess},
\end{align}

where $l$, $p$ are hyperparameters of the length scale and the period, respectively. These kinds of kernels are suitable for data that has features of periodicity. For example, the weather forecasting data and the stock market data. From Fig.~\ref{argu}, the squeezing phase kernel has sharper peaks compared to the exponential sine squared kernel, and may be more suitable for modeling functions with sudden changes.

In addition, for coherent-state encoding~\cite{zhang-2020-PRL}, data can also be encoded into the argument of coherent state that the kernel becomes~(the real part of the inner product), 
\begin{align}
\braket{c,x|c,x'}_c&=\exp\left( -c^2 + c^2 e^{i(x'-x)} \right) \\
\Re{\braket{c,x|c,x'}_c}&= e^{-c^2 + c^2 \cos{(x'-x)}} \cos{(c^2 \sin{(x'-x)})}
\end{align}
The hyperparameter $c$ has the similar function as the length scale $l$ in the exponential sine squared kernel. We call this kernel the {\it coherent phase kernel}. Although there are no parameters for tuning the desired period, this can always be done by rescaling the data with some multiplication factor. Therefore, both phase encoding of squeezing and coherent feature maps are periodic kernels, which can be used to model functions that are periodic, similar to the exponential sine squared kernel.

\subsection{Amplitude encoding}
Another way to encode into squeezed states is to encode the data into the amplitude of squeezing parameter $\phi:x \rightarrow \ket{x,0}$, where the phase is set as zero. There is no hyperparameter in this case since the exponential factor vanishes for every equal value in the squeezing vacuum inner product (Eq.(\ref{spk})) and we set it as zero. The resulting kernel is

\begin{align}
\braket{x,0|x',0}_s=\sqrt{\frac{\sech{x}\sech{x'}}{1-\tanh{x}\tanh{x'}}}
\end{align}

\begin{figure}[h]
	\begin{center}
		\hspace*{0cm}
		\includegraphics[width=0.23\textwidth]{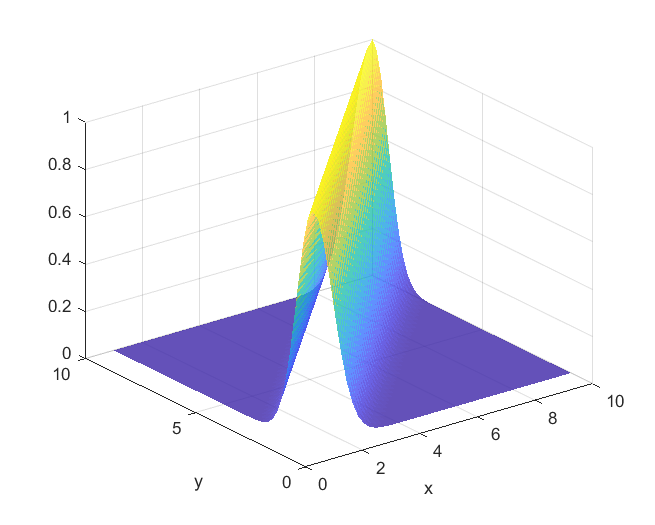}
		\includegraphics[width=0.23\textwidth]{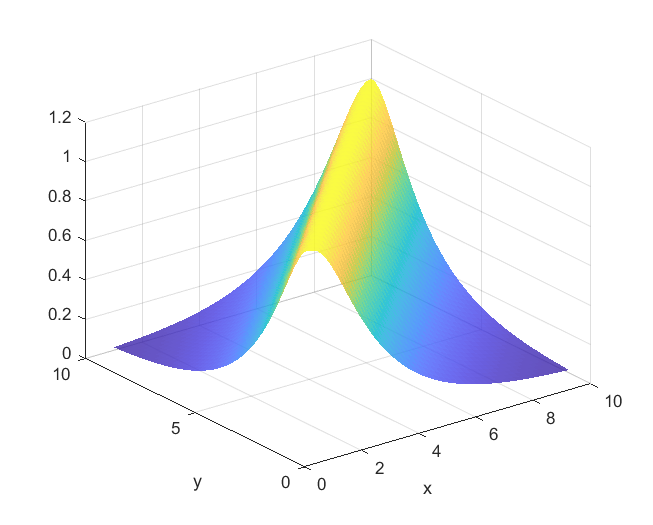}
		\caption{Comparison between Gaussian kernel (left) and squeezing amplitude kernel (right).} \label{amp}
	\end{center}
\end{figure}

We call this {\it squeezing amplitude kernel}. The kernel is symmetric and positive semidefinite, and has a shape similar to the Gaussian kernel, as shown in Fig. \ref{amp}. The inner product formed by the basis encoding of the coherent state into the amplitude has been shown to have the similar functional form of Gaussian kernel~\cite{zhang-2020-PRL}, also with no extra hyperparameter,

\begin{align}
\braket{x,0|x',0}_c=\exp\left(-\frac{1}{2}|x-x'|^2\right)
\end{align}

The squeezing kernel has a larger spread compared to the Gaussian kernel. It can be used to model functions with larger variance, which varies more slowly with respect to the distance from a center point.

\section{Machine learning with quantum kernels}
We now apply quantum kernels with squeezed-state encoding for machine learning. The kernels are obtained through a quantum circuit of evaluating the inner product between two quantum states, and then serve as the input for the support vector machine for training and prediction. We further test those quantum kernels on some standard datasets and compare the performances with those of classical kernels. 

\subsection{Support vector machine with quantum kernels}
There are two main streams of quantum algorithms where the kernel method applies. The first kind is solely based on quantum computers. Famous examples include linear regression~\cite{wiebe_12,zhang_19}, quantum support vector machine~\cite{rebentrost_14}, quantum principal component analysis~\cite{lloyd_14}. The other class is the classical-quantum hybrid approach that a part of computational task is left to classical devices. In this regime, the optimization of the machine learning algorithm is usually outsourced to classical computers. Quantum computers are only required to generate quantum gates that encode data, and then to calculate the inner product between quantum states. This approach is referred to as quantum kernel estimation~\cite{schuld_19}, which is easier to achieve than the first approach as both the circuit depth and the quantum resources required are less. We can use quantum computers to calculate the proposed quantum kernels above, and the result can be used by classical computers for training traditional machine learning models like support vector machines, principal component analysis, clustering, and Gaussian processes. We take the support vector machine as an example.

Firstly, the inner product can be calculated on a quantum computer in two ways. The first is to use the swap test. Taking the squeezing kernel as an example. We need to prepare the quantum state 
\begin{align} \label{eq:eval_kernel}
\ket{\Psi}=\frac{1}{\sqrt{2}}(\ket{0}\ket{z}_s+\ket{1}\ket{z'}_s)
\end{align}
then take a $\sigma_x$ measurement on the ancillary qubit, to measure the probability of getting $+1$. The inner product can be calculated by $2p-1$. Another approach to calculating the inner product is to apply two consecutive squeezing operators $\hat{S}(z)$ then $\hat{S}^\dagger (z')$, and measure the probability that the qumode is in vacuum state $\ket{0}$. This avoids the construction of states that entangle qubit and qumode in Eq. (\ref{eq:eval_kernel}).

Secondly, with kernels at hand, we continue the procedure of machine learning on classical computers. To optimize an SVM classifier, we have to maximize the Lagrangian dual function~\cite{christopher_m_bishop_pattern_2006,trevor_hastie_elements_2009}
\begin{align}
L(\vec{\alpha})=\sum_{i=1}^M \alpha_i - \frac{1}{2}\sum_{i,j=1}^M y_iy_j \alpha_i \alpha_j K(x_i,x_j)
\end{align}
subject to the constraint $\sum_{i=1}^M \alpha_i y_i=0$ and $\alpha_i \geq 0$ for every $i$. Optimized values of $\alpha_i$ and $b$ are
derived from Karush-Kuhn-Tucker (KKT) conditions,
\begin{gather}
\begin{align} y_iy(x_i)&=1 \\
y_i \left(\sum_{j \in S} \alpha_j y_j K(x_j,x_i) +b \right) &= 1 \end{align} \\
b = \frac{1}{N_S} \sum_{i \in S} \left( y_i - \sum_{j \in S} \alpha_j y_j K(x_j,x_i) \right)
\end{gather}
where $S$ denotes the set of support vectors.
Then, prediction can be calculated by 
\begin{align}
y(x)=\sum_{i \in S} \alpha_i y_i K(x_i,x) +b
\end{align}

Therefore, a support vector classifier can be trained after computing the kernel for each pair of data $K(x_j,x_i)$ on the quantum computer, and similar procedures apply to other traditional kernel-based machine learning models. 

\subsection{Results}
\begin{figure}[h]
	\begin{center}
		\hspace*{0cm}
		\includegraphics[width=0.15\textwidth]{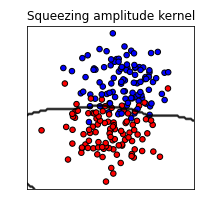}
		\includegraphics[width=0.15\textwidth]{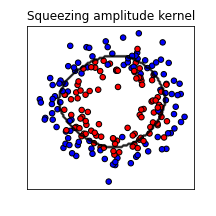}
		\includegraphics[width=0.15\textwidth]{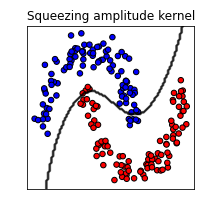}
		\includegraphics[width=0.15\textwidth]{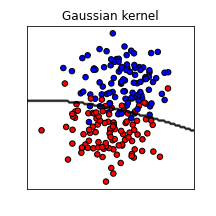}
		\includegraphics[width=0.15\textwidth]{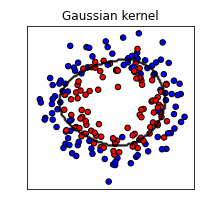}
		\includegraphics[width=0.15\textwidth]{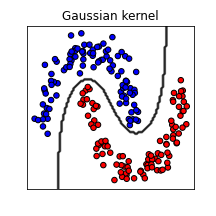}
		\caption{Comparison of SVM training results between squeezing amplitude kernel (first row) and Gaussian kernel (second row). It was trained with Python's scikit-learn SVC classifier using customized kernel. The three standard datasets used are "blobs" (first column), "circles" (second column) and "moons" (third column). $400$ data was generated randomly in each case where the training and validating set have $200$ data each. Only the validating data are shown. The accuracy of each training is measured and shown in Table \ref{tb_acc1}.} \label{svm_sak}
	\end{center}
\end{figure}

\begin{figure}[h]
	\begin{center}
		\hspace*{0cm}
		\includegraphics[width=0.15\textwidth]{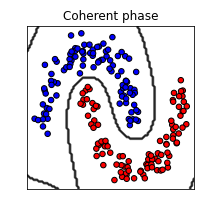}
		\includegraphics[width=0.15\textwidth]{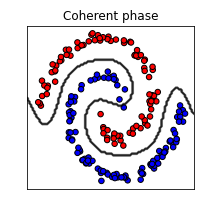}
		\includegraphics[width=0.15\textwidth]{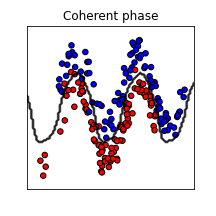}
		\includegraphics[width=0.15\textwidth]{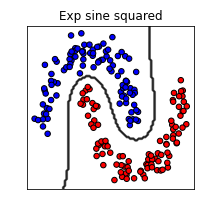}
		\includegraphics[width=0.15\textwidth]{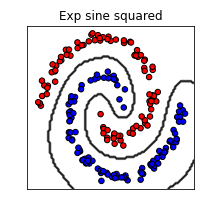}
		\includegraphics[width=0.15\textwidth]{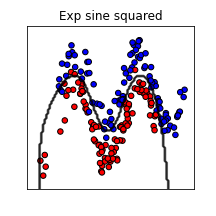}
		\caption{Comparison of SVM training results between coherent phase kernel (first row) and exponential sine squared kernel (second row). It was trained with Python's scikit-learn SVC classifier using customized kernel. The three generated datasets used are "moon" (first column), "spiral" (second column) and "sine" (third column). $400$ data was generated randomly in each case where the training and validating set have $200$ data each. Only the validating data are shown. The accuracy of each training is measured and shown in Table~\ref{tb_acc2}. The hyperparameters are tuned by searching for optimized parameters.} \label{svm_cpk}
	\end{center}
\end{figure}

As an illustration purpose, we test and compare those kernels mentioned in Sec.~\ref{sec:kernels} for supervised learning of classification on some standard datasets from scikit-learn~\cite{scikit-learn}. Those kernels are used to calculate randomly generated sample datasets of each $200$ data, and the SVM model was trained with the kernel. Then, another set of $200$ data is used for validation to calculate the performance score. The results are shown in Fig.~\ref{svm_sak} for comparing squeezing amplitude kernel with the standard Gaussian kernel, and in Fig.~\ref{svm_cpk} for comparing the coherent phase kernel and exponentiated sine squared kernel. Their corresponding performance scores are listed in Table~\ref{tb_acc1} and Table~\ref{tb_acc2}, respectively. The decision boundary of the squeezing amplitude kernel is comparable to the boundary created by the standard Gaussian kernel. The variance of the squeezing amplitude kernel is slightly lower than the Gaussian kernel but still demonstrated the ability to capture the highly nonlinear behaviour of the dataset. On the other hand, the decision boundary created for coherent phase kernel and exponentiated sine squared kernels are different for each case but they shared the common shape in general. These demonstrated that the two kernels will be useful in different data patterns.

\begin{table}[h]
	\begin{tabular}{l|lll}
		& Blobs & Moons & Circles \\
		\cline{1-4}
		Squeezing amplitude & 0.915 & 0.95 & 0.795 \\
		Gaussian & 0.9  & 1.0  & 0.79
	\end{tabular}
	\caption{The accuracy of validation corresponds to  Figure~\ref{svm_sak}.} \label{tb_acc1}
\end{table}

\begin{table}[h!]
	\begin{tabular}{l|lll}
		& Moons & Spiral & Sine \\
		\cline{1-4}
		Coherent phase & 1.0 & 0.995 & 0.96 \\
		Exp sine-squared & 1.0  & 1.0  & 0.95
	\end{tabular}
	\caption{The accuracy of validation corresponds to  Figure~\ref{svm_cpk}.} \label{tb_acc2}
\end{table}
\subsection{Physical implementation}
The whole machine learning procedure only requires the quantum computer at the stage of evaluating quantum kernels, and we briefly discuss its physical implementation on a quantum system.  The inner product has to be calculated on a physical system that allows manipulation of both continuous variables and a discrete-variable qubit as control.  Promising candidates of quantum platforms capable of hybrid variable quantum computing include superconducting circuit and trapped ion quantum computer~\cite{leibfried_03,haffner_08,monroe_13,ortiz-gutierrez17,Zhangjunhua_2018,fluhmann_19}.
We take trapped ion as an example. The motional state of the ion can be used as continuous variables, and various motional states have been realized experimentally~\cite{ortiz-gutierrez17,Zhangjunhua_2018,fluhmann_19}. Remarkably, squeezed vacuum state $\ket{z}_s$ can be created by irradiating two Raman beams with some specific frequency difference. The squeezing parameter will grow exponentially with driving time for the Raman transition that occurs during the process. Moreover, internal levels of an ion can serve as a qubit and its coupling to the motional modes can implement the quantum circuit to generate the state in Eq.~\ref{eq:eval_kernel}
that is central to the evaluation of quantum kernels. 

\section{Conclusions}

We have illustrated the idea of using different continuous-variable quantum states to generate new nonlinear kernels that can be used to learn different data patterns in machine learning. By encoding the data into either amplitude or argument of the parameter of squeezing, different kernels can be obtained. We have applied those quantum kernels with classical support vector machines for classification tasks on some standard datasets, and their performances are comparable or better to typical classical kernels.  Furthermore, we have suggested a quantum platform, the trapped ion quantum computer, which can implement the quantum algorithm involving both discrete-variable qubits and continuous variables. We mention that the squeezing and displacement operator suggested in this article belong to a larger family called Gaussian operations, and general Gaussian-state encoding may be used as quantum kernels for machine learning. Moreover, since Gaussian-state can be characterized by the covariance matrix, a fully exploitation of the infinite dimensionality of continuous variables should refer to non-Gaussian quantum states. The case of non-Gaussian state encoding and its applications as quantum kernels with quantum advantages will be investigated in the future. 

\begin{acknowledgments}
This work was supported by the Key-Area Research and Development Program of GuangDong Province (Grant No. 2019B030330001) and the Collaborative Research Fund of Hong Kong (No. C6005-17G). 
\end{acknowledgments}

\bibliography{squeezing}
\end{document}